%% file: main.tex
\documentclass[
  journal=pasa,
  manuscript=research-paper, 
  year=2025, 2163
  volume=YY,
]{cup-journal}

\usepackage{amsmath}
\usepackage{amssymb,microtype,siunitx,booktabs}

\usepackage{tocloft}
\usepackage{lineno}
\usepackage{booktabs}
\usepackage{setspace}
\usepackage{tipa}
\usepackage{academicons} 
\usepackage{fontawesome5}

\usepackage{amsmath,amsfonts,amssymb}
\usepackage{graphicx}
\usepackage{relsize}

\usepackage{listings}
\usepackage{multirow}

\usepackage{longtable}
\usepackage{acro}
\usepackage{xspace}

\definecolor{color1}{HTML}{1f77b4}
\definecolor{color2}{HTML}{ff7f0e}
\definecolor{color3}{HTML}{2ca02c}
\definecolor{color4}{HTML}{d62728}

\newcommand{\fblue}{\textcolor{color1}{F380M}\xspace}
\newcommand{\fgreen}{\textcolor{color3}{F430M}\xspace}
\newcommand{\fred}{\textcolor{color2}{F480M}\xspace}

\acsetup{
  single    = false,
  list/sort  = true,
  cite/group = true,
  cite/group/cmd = \citealt,
  cite/group/pre = {; \xspace},
  patch/longtable=false,
}

\newcommand{\amigo}{\texttt{amigo}\xspace}

\newcommand\optax{\texttt{optax}\xspace}

\newcommand\dlux{\texttt{$\partial$Lux}\xspace}
\newcommand\dorito{\texttt{dorito}\xspace}
\newcommand\ami{AMI\xspace}

\input{acronyms}
\usepackage[utf8]{inputenc}

\usepackage{multirow}
\usepackage{colortbl}
\usepackage{arydshln}

\usepackage{hyperref}
\hypersetup{
    colorlinks,
    linkcolor={color1},
    citecolor={color1},
    urlcolor={color1}
}
\usepackage{cleveref}
\AddToHook{cmd/appendix/before}{\crefalias{section}{appendix}}

\usepackage{siunitx}
\DeclareSIUnit\year{yr}
\DeclareSIUnit\parsec{pc}
\DeclareSIUnit\arcsecond{as}
\usepackage{gensymb}

\usepackage{natbib}
\usepackage{subcaption}

\title{Image reconstruction with the JWST Interferometer}

\author{\href{https://orcid.org/0009-0003-5950-4828}{Max~Charles}}
\affiliation{Sydney Institute for Astronomy, School of Physics, University of Sydney, Camperdown, NSW 2006, Australia}
\email[Max Charles]{\href{mailto:max.charles@sydney.edu.au}{max.charles@sydney.edu.au}}
 
\author{\href{https://orcid.org/0000-0002-1015-9029}{Louis~Desdoigts}}
\affiliation{Sydney Institute for Astronomy, School of Physics, University of Sydney, Camperdown, NSW 2006, Australia}
\alsoaffiliation{Leiden Observatory, Niels Bohrweg 2, Leiden 2300RA, The Netherlands}

\author{\href{https://orcid.org/0000-0003-2595-9114}{Benjamin~Pope}}
\affiliation{School of Mathematical \& Physical Sciences, Macquarie University, 12 Wally's Walk, Macquarie Park, NSW 2113, Australia}
\alsoaffiliation{School of Mathematics \& Physics, University of Queensland, St Lucia, QLD 4072, Australia}

\author{\href{https://orcid.org/0000-0001-7026-6291}{Peter~Tuthill}}
\affiliation{Sydney Institute for Astronomy, School of Physics, University of Sydney, Camperdown, NSW 2006, Australia}

\author{\href{https://orcid.org/0000-0001-9582-4261}{Dori~Blakely}}
\affiliation{Department of Physics and Astronomy, University of Victoria, 3800 Finnerty Road, Elliot Building, Victoria, BC V8P 5C2, Canada}
\alsoaffiliation{NRC Herzberg Astronomy and Astrophysics, 5071 West Saanich Road, Victoria, BC V9E 2E7, Canada}

\author{\href{https://orcid.org/0000-0002-6773-459X}{Doug~Johnstone}}
\affiliation{Department of Physics and Astronomy, University of Victoria, 3800 Finnerty Road, Elliot Building, Victoria, BC V8P 5C2, Canada}
\alsoaffiliation{NRC Herzberg Astronomy and Astrophysics, 5071 West Saanich Road, Victoria, BC V9E 2E7, Canada}

\author{\href{https://orcid.org/0000-0003-2259-3911}{Shrishmoy~Ray}}
\affiliation{School of Mathematical \& Physical Sciences, Macquarie University, 12 Wally's Walk, Macquarie Park, NSW 2113, Australia}
\alsoaffiliation{School of Mathematics \& Physics, University of Queensland, St Lucia, QLD 4072, Australia}

\author{\href{https://orcid.org/0000-0002-5956-851X}{K.~E.~Saavik Ford}}
\affiliation{Center for Computational Astrophysics, Flatiron Institute, 
162 5th Ave, New York, NY 10010, USA}
\alsoaffiliation{Department of Astrophysics, American Museum of Natural History, New York, NY 10024, USA}
\alsoaffiliation{Department of Science, Borough of Manhattan Community College, City University of New York, New York, NY 10007, USA}

\author{\href{https://orcid.org/0000-0002-0786-7307}{Barry McKernan}}
\affiliation{Center for Computational Astrophysics, Flatiron Institute, 
162 5th Ave, New York, NY 10010, USA}
\alsoaffiliation{Department of Astrophysics, American Museum of Natural History, New York, NY 10024, USA}
\alsoaffiliation{Department of Science, Borough of Manhattan Community College, City University of New York, New York, NY 10007, USA}

\author{\href{https://orcid.org/0000-0003-1251-4124}{Anand~Sivaramakrishnan}}
\affiliation{Space Telescope Science Institute, 3700 San Martin Drive, Baltimore, MD 21218, USA}
\alsoaffiliation{Astrophysics Department, American Museum of Natural History, 79th ST at CPW, New York, NY, 10024, USA}
\alsoaffiliation{Department of Physics and Astronomy, Johns Hopkins University, 3701 San Martin Drive, Baltimore, MD 21218, USA}

\received {20 October 2025}
\revised  {26 February 2026}
\accepted {12 March 2026}
\published{13 October 2025}


\begin{document}

\begin{abstract}
Flying on board the \textit{James Webb Space Telescope} (JWST) above Earth's turbulent atmosphere, the \ac{ami} on the NIRISS instrument is the highest-resolution infrared interferometer ever placed in space. However, its performance was found to be limited by non-linear detector systematics, particularly charge migration --- or the Brighter-Fatter Effect. Conventional interferometric Fourier observables are degraded by non-linear transformations in the image plane, with the consequence that the inner working angle and contrast limits of AMI were seriously compromised. Building on the end-to-end differentiable model \& calibration code \href{http://github.com/louisdesdoigts/amigo}{\amigo}, we here present a regularised maximum-likelihood image reconstruction framework \href{https://github.com/maxecharles/dorito}{\dorito} which can deconvolve AMI images either in the image plane or from calibrated Fourier observables, achieving high angular resolution and contrast over a wider field of view than conventional interferometric limits. This modular code by default includes regularisation by maximum entropy, and total variation defined with $l_1$ or $l_2$ metrics. We present imaging results from \dorito for three benchmark imaging datasets: the volcanoes of Jupiter's moon Io, the colliding-wind binary dust nebula WR~137 and the archetypal Seyfert 2 active galactic nucleus NGC 1068. In all three cases we recover images consistent with the literature at diffraction-limited resolutions.
The performance, limitations, and future opportunities enabled by \amigo for AMI imaging (and beyond) are discussed.
\end{abstract}

\section{Introduction}

Following its successful launch, the \ac{jwst} immediately established itself as the world's leading observatory for infrared astronomy. Included in its suite of revolutionary instruments is the \ac{niriss}, hosting the Aperture Masking Interferometer \citep[AMI;][]{Sivaramakrishnan_2023} as one of its four observing modes. The aperture mask, or \ac{nrm}, is a modest titanium disc small enough to fit into the \ac{niriss} pupil wheel perforated with seven holes in a non-redundant pattern to enable interferometric observations.

The venerable technique of sparse aperture masking \citep[or Fizeau interferometry;][]{fizeau1868} blocks much of the telescope aperture, passing light only through a pattern of holes. Pairs of these create interference fringes in the point spread function, offering established calibration advantages compared to full-pupil imaging. If the array is non-redundant \citep{Tuthill2012}, with no two holes separated by the same vector, they form unique Fourier amplitudes \citep{Twiss1960} and (importantly) closure phases \citep{closure_phase,optical_cps} that are self-calibrating with respect to the effects of optical aberrations from the turbulent atmosphere and (relevant to \ac{jwst}) imperfect/variable surface figure of mirrors. Closure phases can therefore provide extremely accurate constraints on faint companions and asymmetric structure in images, even in the presence of systematic optical errors.

Prior to launch, \ami was expected to debut as the most scientifically capable space interferometer to date, boasting unparalleled angular resolving power and contrast sensitivity \citep{Sivaramakrishnan_2010,2012_sivaramakrishnan,2020Soulain, 2022_tuthill}. 
A number of pipelines have been developed to analyse the \ami data: \texttt{AMICAL} \citep{amical}, \texttt{SamPy} \citep{sampy}, and \texttt{SAMPip} \citep{sampip} build on legacy Fizeau interferometry codes using the \ac{fft} to sample visibilities in the Fourier plane. In contrast, the forward-modelling code \texttt{ImPlaneIA} fits to the interferogram directly in the image plane \citep{ImPlaneIA}, and extracts visibility information from this fringe fit. 

However, it turned out that all codes reached sensitivity limits in faint companion searches that were severely hindered by non-linear charge migration in \ac{niriss}' HAWAII-2RG \ac{cmos} detector \citep{2024_Sallum, Ray2025}. Charge diffuses from brighter pixels into neighbouring pixels, resulting in bright features appearing broader or fatter; hence this charge migration has been dubbed the \ac{bfe}. \ami's image plane is barely Nyquist-sampled by the detector, which exacerbates its sensitivity to charge migration. Crucially, the non-linearity of the \ac{bfe} cannot be calibrated in the Fourier plane alone. Repeated measurements sampling up-the-ramp has allowed for data selection mitigating the impact of \ac{bfe}, yielding precise measurements of the protoplanetary disk and two companions in the PDS~70 system \citep{doripds70}.
However, the lack of a general solution for \ac{bfe} calibration has severely hindered science productivity of programs requiring precise calibration at diffraction limited scales.

In a companion paper \citep{Desdoigts2025}, we present a new pipeline, \href{http://github.com/louisdesdoigts/amigo}{\amigo} (AMI Generative Optics), a \texttt{Python} package for end-to-end differentiable forward-modelling of \ami. This connects a differentiable optical model written in \texttt{Jax} \citep{jax} and \dlux \citep{dLuxI,dLuxII} with a neural network ``effective detector model'' that models the \ac{bfe} and other detector effects. These are simultaneously trained on an ensemble of calibration datasets, to arrive at a base model with fixed, fiducial optical and detector parameters that reflect the average state of the instrument. This base model can then be applied to retrieve astrophysical information from \ami data through forward-modelling and Bayesian inference with optimisation or sampling. Here we allow the surface figures of mirrors and the parameters of astrophysical scenes to vary while holding the mask metrology and electronic properties fixed. By directly modelling pixel-level data as they accumulate up-the-ramp, \amigo is able to sidestep the challenges of Fourier-plane calibration and improve upon the performance of the \hyperlink{https://jwst-docs.stsci.edu/jwst-science-calibration-pipeline}{\ac{jwst} pipeline} significantly. In the \amigo companion paper, we apply this new pipeline to the recovery of faint companions at the diffraction \textit{and} photon noise limit in the HD~206893 system \citep{hd206893_orbit} and in the binary AB~Dor~AC \citep{Azulay2017}. 

In essence, \amigo enables accurate modelling of the \ami \ac{psf} and astrophysical scenes composed of unresolved point sources.
In this paper, we extend the capability of amigo to include arbitrary resolved scenes, implemented in the \texttt{Python} package \href{https://github.com/maxecharles/dorito}{\dorito} (Differentiable Optics for Reconstructing Images from Telescope Observations).
We document and apply \ac{rml} image reconstruction methods fitting to either the image plane or interferometric observables. 
We explore several different methods for image reconstruction, finding that in most cases modelling the source distribution directly to the image plane rather than to visibilities or closure quantities yields the best performance.

\begin{figure*}[ht!]
\begin{center}
\includegraphics[width=1\textwidth]{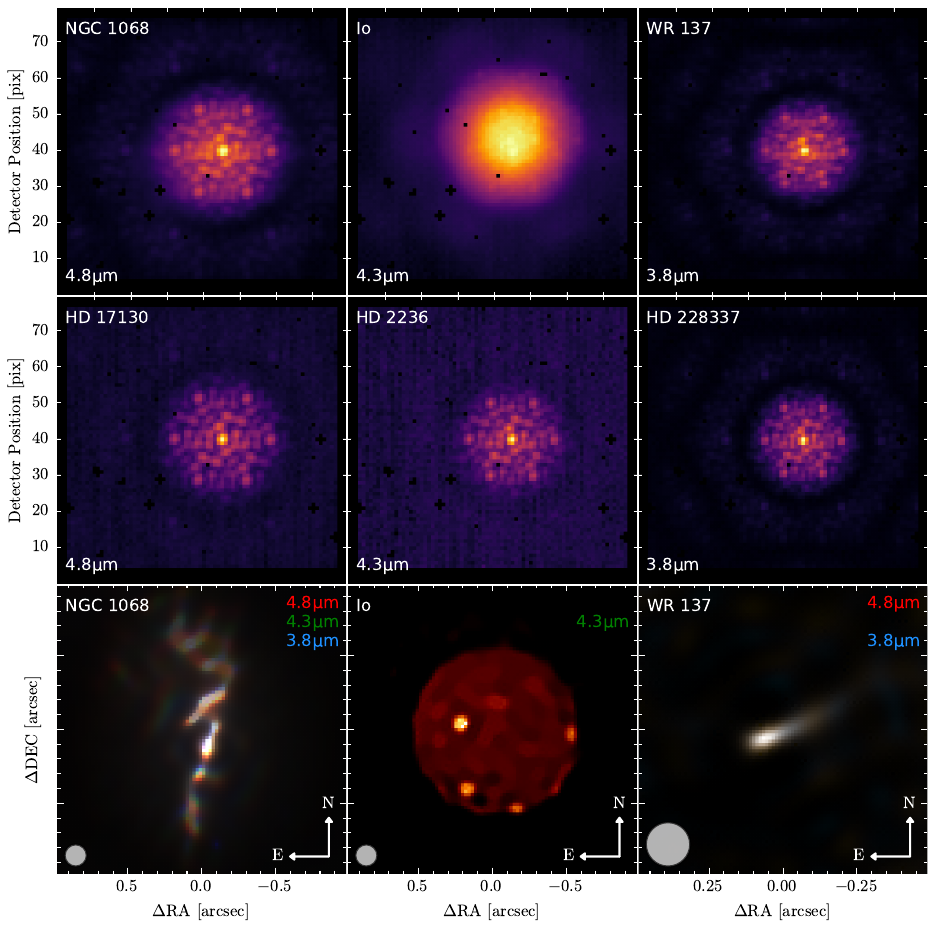}
\end{center}
\caption 
{Interferograms of the science targets, calibrators, and deconvolved images.
\textit{Top}: Images of the interferograms of science targets NGC~1068, Io, and WR~137. These images encode the slope of the ramp integrations, i.e. final group subtract second-to-final group. They contain noticeably less high frequency power by comparison to the \textit{middle} row of \ac{psf} calibrators corresponding to the above sources. Bad pixels not used in the fit are set to black.
\textit{Bottom}: \ac{rml} image reconstructions of all three targets. The beam size circle in the lower left of each panel indicates $\lambda/2D$. The colour tables for each image is a power law stretch ($\gamma=0.4, 0.8, 0.3$ respectively for each target left to right). NGC~1068 is clipped to $25\%$ of its peak brightness in the \fred filter. Note WR~137 is displayed on a factor of two finer pixel scale than the other images due to its small angular size. NGC~1068 and WR~137 are both presented as RGB false-colour images weighted by the recovered flux values in each filter. However as WR~137 was only observed in the \fred and \fblue filters, the green channel was assigned the average of the other two channels. The predominant white hue of these two images is a sign that the same structure is independently recovered in all three bands, a sign of successful deconvolution of a source without strongly wavelength-dependent features. The Io image is a single exposure image, ignoring the other epochs of data.} 
\label{fig:data}
\end{figure*}

We present imaging results from three public \ami datasets: the active galaxy NGC~1068, Jupiter's moon Io \citep[previously published by][]{Sanchez-Bermudez2025}, and the colliding-wind binary \acl{wr137} \citep[\ac{wr137}; previously published by][]{2024ApJ_RyanLau}. These each serve to illustrate important points about the performance of imaging with \ami. The science and calibrator data are presented alongside their reconstructed images in Figure~\ref{fig:data}.

NGC~1068 is a well-known \ac{agn} whose inner regions have recently been imaged in the mid-infrared \citep{Isbell25} using the \ac{lbti}. Here, we recover a very closely matching structure in all three bands of \ami.

The innermost Galilean moon Io has a surface pocked with violent volcanism, visible from Earth in Keck adaptive optics images \citep{Marchis2002,deKleer2019} and close up from NASA Juno flyby observations \citep{Mura2020}. An independent analysis of the \ami images by \citet{Sanchez-Bermudez2025} using a neural network reveals bright spots at the locations of known volcanoes; we reproduce this result but at higher resolution, seeing a crisp edge to the disk of Io and well-resolved spots rotating over the hour-long observation.

Colliding-wind binaries like \ac{wr137} \citep{WR1371985,WR1372001} produce spiral plumes of dust that were some of the first images to be resolved by aperture-masking interferometry \citep{Tuthill1999}. In \ami observations of \ac{wr137}, \citet{2024ApJ_RyanLau} show that its dusty environment is clearly resolved into a streak lined up with the expected elongation of the plume. An interesting and unfortunate quirk of this dataset is that, during the sequence of calibrator and science target observations with \ami, there was a `tilt event' \citep{Schlawin2023,Lajoie2023} in which one of the mirror segments abruptly moved, so that the wavefront encountered by the \ac{psf} reference calibrator differs from that of the science target. While in general we show in this paper that closure invariants perform poorly for anchoring image reconstructions with \ami compared to image plane modelling, in this case we find that they significantly improve the deconvolution with significant wavefront error present.

\subsection{Differentiable forward-modelling of the instrument}

This work employs a full end-to-end differentiable forward-model approach to image reconstruction. This section motivates this choice over alternatives such as an inverse model approach.

An instrument, or sensing system, delivers measurements $\mathbf{y}$ which encode some desired signal $\boldsymbol{\theta}$. The measurements and signal are related by some function $f$ such that 

\begin{equation}
\mathbf{y} = f(\boldsymbol{\theta}) + \boldsymbol{\epsilon},
\end{equation}

\noindent where $\boldsymbol{\epsilon}$ is some non-deterministic, independent, zero-mean noise term.
Therefore an inverse problem must be solved to recover $\boldsymbol{\theta}$. The obvious approach is to construct an inverse function such that $\boldsymbol{\theta} = f^{-1}(\mathbf{y})$. However experience shows that this quickly proves arduous for complex non-linear problems and is particularly problematic when the effects of non-deterministic noise processes need to be compensated. Crucially, for the ill-posed problem of image reconstruction, an inverse function cannot be uniquely defined where $f$ is not a one-to-one mapping.

The forward-model approach instead simply constructs the function $f$ --- an end-to-end forward-model of the instrument --- to best emulate the deterministic physics at play.
This is effective in a Bayesian framework as Bayesian methods can be used to infer information about the signal $\boldsymbol{\theta}$ from the data $\mathbf{y}$. The optimisation problem in \Cref{eq:bayes1} can be re-expressed in terms of the forward-model as

\begin{equation}
\boldsymbol{\hat{\theta}} = \arg\max_{\ \boldsymbol{\theta}} \left[ \underbrace{\mathcal{L}\left( \mathbf{y} | f(\boldsymbol{\theta}) \right)}_{\text{log-likelihood}} + \underbrace{\Pi\left(f(\boldsymbol{\theta})\right)}_{\text{log-prior}} \right].
\label{eq:bayes2}
\end{equation}

\noindent Here $\boldsymbol{\theta}$ are the model parameters, which in the case of the \amigo model includes the image parameters $\mathbf{c}$, other source parameters (such as detector position, flux), and optical/detector instrumental parameters (such as mirror aberrations, pixel gains, and charge migration). During optimisation, it is then possible to jointly solve for the instrumental state as well as the source parameters.

As the model complexity increases, so do the computational demands of optimisation.
It is advantageous to build the forward-model in a differentiable coding language such as \texttt{Jax} \citep{jax}, \texttt{PyTorch} \citep{pytorch}, or \texttt{TensorFlow} \citep{tensorflow}, as they leverage \ac{autodiff} to facilitate the use of gradient-based optimisation methods.
These include straightforward optimisation engines such as gradient descent \citep{deepmind2020jax}, but also encompass modern efficient gradient-based \ac{mcmc} algorithms for error estimation such as \ac{hmc} and \ac{nuts}.
These languages also allow for \ac{gpu}-acceleration, drastically increasing the tractability of training models with millions or billions of parameters on relatively accessible hardware. 


Our \ac{rml} deconvolution code \dorito is inspired by \texttt{MPoL}, an \ac{rml} imaging framework written in \texttt{pytorch} built for reconstructing radio/sub-mm interferometric data \citep{mpol}. \texttt{MPoL} has been used to show the effectiveness of RML techniques improving image fidelity on ALMA continuum observations of protoplanetary disks \citep{mpolalma}.

\subsection{Data Reduction and Wavefront Calibration}

Optical interferometers must perform physical beam combination to interfere the light: for Fizeau interferometry (as performed by AMI) we refer to the fringe pattern on the detector as the interferogram. This pattern is the convolution of the \ac{psf} of the telescope with the source intensity distribution. The \ac{psf} is usually modelled as the Fourier (or Fresnel) transform of the incoming wavefront, which for high fidelity recovery requires the inclusion of phase distortions --- whether from the turbulent atmosphere or, even in space, from optical path delays due to uneven or misaligned mirror surfaces.

Aperture synthesis is still possible in the presence of wavefront phase distortions by transforming the measured complex visibilities into data products that are less vulnerable to phase error. This includes closure quantities such as closure phases \citep{closure_phase} or kernel phases \citep{kernel_phase}, or extraction of squared visibilities from the power spectrum (which simply discards phase information).

In the workflow presented here, the first step of any data processing is to infer the wavefront. We load a pre-saved \ami base model trained as described in the \amigo companion paper, and hold the parameters describing mask metrology and the effective detector model fixed. We take the \ac{psf} reference star observed along the science target, and fit for the wavefront across each aperture with the trained \amigo base model. This is parametrised as a set of Zernike polynomials describing phase deviations across each hole of the aperture mask. We then hold these parameters fixed and analyse the science target, either directly modelling the source distribution as an array (Method 1, \Cref{sec:method1}) or first extracting and appropriately calibrating complex visibilities (Method 2, \Cref{sec:method2}). 
 
\subsection{Image reconstruction}

While in the \amigo companion paper we detect point source companions at high contrast, in this paper we aim to reconstruct complex resolved scenes. 
Image reconstruction is the process of solving the inverse problem of finding the image $I$ which produced measured data $\mathbf{y}$. 
Image reconstruction has a long history in both radio and optical interferometry \citep{baron2016image,ThiebautPrinciples}, where a limited subset of the Fourier modes in an image are measured.
In general, where information is lost to blurring or filtering (as in interferometry), the inverse problem is ill-posed; that is, there is no unique solution $I$ which could produce $\mathbf{y}$ because many possible solutions fit the data equally well. Over the last century, this has spurred the invention of many algorithmic approaches to find the best of the many possible solutions.

The simplest of these methods is direct aperture synthesis by linearly adding the interference patterns, where an array of telescopes (or telescope sub-apertures, in the case of AMI) are cophased to achieve an angular resolution equivalent to that of a single aperture the size of the longest baseline. If the sub-apertures are arranged in a non-redundant pattern such that no two are separated by the same vector, each interferometric baseline is unique and each telescope pairing will sample a different region of the $u$-$v$ plane. Assuming the Fourier amplitudes are zero where the $u$-$v$ plane has not been sampled, one can then inverse transform from the $u$-$v$ plane back to the image plane to obtain a reconstructed image. In radio astronomy, this resultant image is referred to as the ``dirty image''. 
Linear aperture synthesis alone can lead to images plagued by artefacts and sidelobes due to incomplete sampling of the Fourier plane, as is indicative by the ``dirty image'' nomenclature. This can lead to the reconstruction of non-physical structure, however the dirty image serves as a starting point for more advanced methods.

One such older approach is the CLEAN method, which was designed to tackle interferometric data with sparse and irregular sampling of the $u$-$v$ plane \citep{clean, clean_schwarz, clean_clark, clean_cornwell}.
Commonly employed in radio astronomy, the CLEAN method is an iterative deconvolution algorithm that models the observed image as a collection of point sources and progressively subtracts copies of the instrument's \ac{psf} (or dirty beam) to reconstruct a ``cleaned'' image.
However, CLEAN is sensitive to phase errors and implicitly assumes the scene is composed of point sources, so it is not optimal for reconstructing extended structure. 
Despite this, CLEAN is still widely and successfully used to reconstruct resolved scenes for longer wavelength observations such as ALMA \citep{modernclean} where $u$-$v$ coverage is extensive and phase error is less destructive.

More recent Bayesian-based methods are better equipped to deal with phase errors and do not impose restrictive assumptions on the image. Within the Bayesian framework, the problem is re-framed as an optimisation or sampling problem inferring the parameters of a forward-model. One finds an image $I$ which maximises the posterior probability, given the likelihood of that image matching the data and given prior assumptions. Mathematically, we wish to solve

\begin{equation}
\label{eq:bayes1}
\hat{I} = \arg\max_{I} \left[ \underbrace{\mathcal{L}\left( \mathbf{y} | I \right)}_{\text{log-likelihood}} + \underbrace{\Pi\left(I\right)}_{\text{log-prior}} \right],
\end{equation}

\noindent where $\mathbf{y}$ is our observed data product (e.g. visibilities, closure phases) and $\hat{I}$ is the maximum a posteriori image. Different competing Bayesian approaches are generally distinguished by their implementation of the Bayesian prior, or we can step outside the Bayesian framework and add regularisers which may encode prior information but are not themselves admissible prior distributions.

\subsection{Image basis}
An implicit way to impose prior knowledge is by the parametrisation of the image \citep{thiebaut_mira_2008}. Consider a basis of $n$ functions (also called a dictionary) used to represent an image:

\begin{equation}
z = \{ z_j : \Omega \to \mathbb{R} \mid j = 1, 2, \ldots, n \},
\end{equation}

\noindent with image coordinates $\alpha, \delta$ in the field of view $\Omega$, expressed in angular units. A continuous representation of the image $I$ is then written as a linear combination of these basis functions:
\begin{equation}
I(\alpha, \delta) = \sum_{i=1}^{n} c_i z_i(\alpha, \delta), \quad \quad (\alpha, \delta) \in\Omega.
\end{equation}

\noindent The coefficients $\mathbf{c}=(c_1,\ldots,c_n)$ are the image parameters to recover in the reconstruction process.
One could build a parametric model of the astrophysical scene from a set of astrophysical parameters, restricting the resulting ``image'' to be of that form.
An example of this is the \ac{ami} parametric model image of PDS~70 reconstructed by \citet{doripds70}.

On the other hand, one could make no assumptions about the scene --- the most obvious parametrisation here is the case where the image parameters are simply the pixel values. The pixel basis, sometimes referred to as the impulsion basis \citep{baron2016image}, is expressed in the above notation as a set of Kronecker delta functions

\begin{equation}
\label{eq:pixelbasis}
z_j(\alpha, d) = \delta_{\alpha_j \alpha} \delta_{d_j d},
\end{equation}

\noindent where each function $z_j$ is an impulse activated at the corresponding pixel centre $(\alpha_j,d_j)$ --- here the latter coordinate is written as $d$ to disambiguate from the Kronecker delta function, but is written as $\delta$ elsewhere. 

Another option is a wavelet basis. This is a natural choice as most physical images, including astrophysical scenes, are sparse in a wavelet basis \citep{OG_wavelets}. Reconstruction in a wavelet basis with $l_1$ regularisation on the wavelet coefficients (see \Cref{sec:lp}) will confine the space of possible solutions to those with more physical characteristics. The \texttt{Mayonnaise} image reconstruction pipeline \citep{mayonnaise} --- developed to reconstruct near-infrared circumstellar disk and exoplanet images in the context of high contrast angular differential imaging --- successfully implements a shearlet basis, which further extends the concept of wavelets to an isotropic basis.

All images presented in this paper are reconstructed in a pixel basis, as described by \Cref{eq:pixelbasis}. Hence the recovered image parameters $\textbf{c}$ are simply the pixel values of the image, and there is no significant prior constraint on the resultant images due to the choice of image basis. We thus turn to image regularisation to apply prior constraint to the images.

\subsection{Regularisation}

The explicit way to incorporate prior knowledge is through the choice of log-prior function $\Pi$ in \Cref{eq:bayes1}.
A common implementation of $\Pi$ is with a regularisation function $\mathcal{R}$, which provides a measure of the discrepancy between a proposed image and its expected characteristics \citep{2011RenardRegul}.
Regularisation directly addresses the ill-posed nature of the inverse problem, and is especially necessary when working with sparse data. It can help to ``paint in'' the regions of the $u$-$v$ plane which are not constrained directly by the data, as opposed to assuming they are zero as in simple aperture synthesis.

Depending on the choice of $\mathcal{R}$, regularisation penalises a candidate image for not satisfying certain desired image properties expected of the scene; e.g. smoothness, soft edges, or sparsity. This is crucial in filtering out non-physical solutions from the solution space which would satisfy the data penalty alone. Here, we optimise not just a log-likelihood, but a more general loss function.
The contribution of the regularisation function to the loss function is weighted by a manually-set hyperparameter $\lambda$

\begin{equation}
\text{Loss} = -\mathcal{L}(\mathbf{y}|I) + \Pi(\mathcal{I}) =-\mathcal{L}(\mathbf{y}|I) + \lambda \mathcal{R}(\mathcal{I}),
\end{equation}

\noindent where $\mathcal{I}$ is the image $I$ integrated onto a discrete pixel grid, with pixel indices $(i, j)$.
This is the backbone of \ac{rml} methods and is leveraged by many existing interferometric image reconstruction packages such as \texttt{bsmem} \citep{BSMEM}, \texttt{Mira} \citep{thiebaut_mira_2008}, \texttt{Squeeze} \citep{squeeze}, \texttt{Painter} \citep{painter}, \texttt{Wisard} \citep{wisard}, 
and \texttt{MPoL} \citep{mpol}, and was used in the reconstruction of the black hole images from the \citet{EHT1, EHT4}.

In order for a regularisation function $\mathcal{R}$ to qualify as a true Bayesian prior distribution, it must be a valid probability distribution.
Regularisers which do not meet this condition (such as maximum entropy) are not compatible with true Bayesian methods that characterise the image posterior distribution such as \ac{hmc} \citep{1982_MEM, baron2016image, 2017_Sallum}. However these non-Bayesian regularisers are still useful if the method only seeks a maximum a posteriori solution, such as gradient descent.

\subsubsection{Common Image Regularisers}
The choice of both $\mathcal{R}$ and $\lambda$ depends on judgement. Listed below are some common choices for the regularisation function $\mathcal{R}$. Note that when reconstructing in a pixel basis, the regularisation is being applied to the pixel values themselves, which is the case we will discuss. Different image bases have significantly different responses to the same regulariser.

\begin{figure*}[ht!]
\begin{center}
\includegraphics[width=1\textwidth]{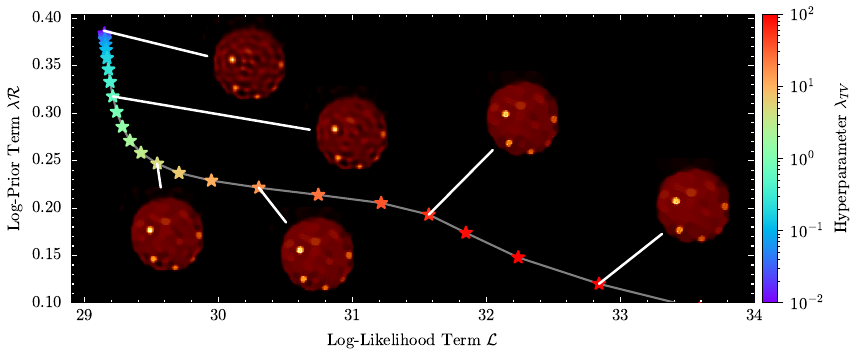}
\end{center}
\caption 
{L-curve diagram used to select the optimal regularisation parameter $\lambda_{\text{TV}}$ for images of Jupiter's moon Io from AMI data. Each point on the curve is the balance between regulariser term $\mathcal{R}$ and likelihood term $\mathcal{L}$ of a converged image reconstruction for a different regularisation hyperparameter value $\lambda_{\text{TV}}$. Shown are several reconstructed images corresponding to different points along the curve. The effects of TV regularisation on the image can be seen to strengthen with increasing $\lambda_{\text{TV}}$ (as regularisation increases, ringing artefacts reduce and plateaux of uniform flux with sharp edges form). The optimal value for $\lambda_{\text{TV}}$ is selected from the region of the elbow in the curve (the third image). L-curves for NGC~1068 and WR~137 are included in \ref{appendix:lcurves}.} 
\label{fig:io-Lcurve}
\end{figure*}

\subsubsection*{Maximum Entropy}
\vspace{0.4cm}
\begin{equation}
\mathcal{R}_{\text{MaxEnt}}(\mathcal{I}) := -\sum_{i,j} \mathcal{I}_{i,j}\ln{\mathcal{I}_{i,j}}
\end{equation}
A maximum entropy or ``MaxEnt'' regulariser seeks to maximise the total Shannon entropy of the image \citep{Shannon}. An image with the maximum entropy has the minimum amount of configurational information, and thus there must be evidence in the data to justify the presence of any structure \citep{MEM}.
A pioneering implementation of maximum entropy methods to optical interferometry was the code \texttt{bsmem} \citep{BSMEM}, with the method accumulating a proven track record.

\subsubsection*{$l_p$ norms}
\label{sec:lp}
\begin{equation}
\mathcal{R}_{l_p}(\mathcal{I}) := \left[\sum_{i,j}\left|{\mathcal{I}_{i,j}}\right|^p\right]^{1/p},\ \ \ p \geq 1
\end{equation}
$l_p$ norms generalise the idea of distance to depend on different powers of the coordinates, where if $p=2$ we recover the familiar Euclidean distance. If we introduce a regularisation term to minimise the $l_2$ norm of the coefficients of a model, this is often called ridge regression \citep{ridge1, ridge2} or automatic relevance determination, which can be seen as a normal prior on the model coefficients centred at zero that suppresses large values of the coefficients \citep{2011RenardRegul}.

Regularisation with an $l_1$ norm, which in linear models is called \ac{lasso} regression, can be interpreted probabilistically as a zero-centred exponential prior on the absolute value of the coefficients. This has the effect of promoting sparsity in the parameters; i.e. if possible, many or most of the parameter values will go to zero, with a few nonzero parameter values singled out as relevant.
This is an important fundamental concept in the field of compressed sensing \citep{daycare, cs_ieee}. In a pixel basis, $l_1$ encourages a dark field populated by a few bright pixels, which is not useful for the extended structure of our three targets here; but many images are sparse in an appropriate basis, such as a wavelet basis, and $l_1$ regularisation on these coefficients may be useful in future work. 

\subsubsection*{Quadratic Variation (QV)}


\vspace{0.4cm}
\begin{equation}
\mathcal{R}_{\text{QV}}(\mathcal{I}) := \sum_{i,j} (\mathcal{I}_{i+1,j} - \mathcal{I}_{i,j})^2 + (\mathcal{I}_{i,j+1} - \mathcal{I}_{i,j})^2
\end{equation}
Quadratic variation, also known as total squared variation or quadratic smoothness, is the $l_2$ norm on the spatial gradient of the image. It is computed over the finite difference gradient of the image as above. As QV penalises large gradient variations, it discourages neighbouring pixels to differ in value and thus smooths the image.

\subsubsection*{Total Variation (TV)}
\vspace{0.4cm}
\begin{equation}
\mathcal{R}_{\text{TV}}(\mathcal{I}) := \sum_{i,j} \left|\mathcal{I}_{i+1,j} - \mathcal{I}_{i,j}\right| + \left|\mathcal{I}_{i,j+1} - \mathcal{I}_{i,j}\right|
\end{equation}
Total variation is the $l_1$ norm on the spatial gradient of the image. The expression for this is written above, however the absolute value function is not differentiable about zero. Consequently, gradient-based parameter optimisation methods such as gradient descent may exhibit unusual behaviour about this point due to numerical instabilities. It is thus preferable to instead use the following:

\begin{equation}
\mathcal{R}_{\text{TV}}(\mathcal{I}) := \sum_{i,j} \sqrt{(\mathcal{I}_{i+1,j} - \mathcal{I}_{i,j})^2 + (\mathcal{I}_{i,j+1} - \mathcal{I}_{i,j})^2},
\end{equation}

\noindent as this is isotropic, differentiable at zero, and approaches the true $l_1$ norm with increasing parameter values.

Like QV, TV attempts to reduce variation in the image. However, it is made distinct from QV by preferring sparse spikes in gradient between flat regions as opposed to a gentle slope. TV favours images with plateaux of flux, and consequently edges are well-preserved \citep{tv}. Over-regularisation with total variation will result in an image with cartoonish features.

\subsubsection{Hyperparameter selection}

\noindent To choose the hyperparameter $\lambda$ is to walk the line between under-regularisation (where the solution matches the data well but may be unphysical) and over-regularisation (where the solution no longer matches the data).
Visual inspection is almost always sufficient, as the negative effects of under- and over-regularisation are not subtle and there usually exists a wide interval of $\lambda$-values which strike a balance \citep{ThiebautPrinciples}.

A less heuristic approach is the L-curve method, which involves plotting the log-likelihood term $\mathcal{L}$ against the log-prior term $\Pi$ of an optimised image for a range of $\lambda$-values \citep{Hansen1992_lcurve}. In a well-behaved loss space, varying $\lambda$ should trace out a curve in the shape of an `L', and the optimal value of $\lambda$ corresponds to the elbow of the curve. This ensures the largest effect of the regulariser without incurring an excessive data penalty. An example L-curve used for regularisation hyperparameter selection in this work is shown in \Cref{fig:io-Lcurve}, for the case of Jupiter's moon Io. Additionally, L-curves for both NGC~1068 and WR~137 are included in \ref{appendix:lcurves}.

\section{Image Reconstruction Methods}\label{sec:methods}

In this paper, we employ two methods for image reconstruction of \ami data. Both methods use gradient descent on a pixel basis to obtain a maximum a posteriori image, with a Gaussian log-likelihood function.

To avoid covariance with the flux parameter, the source distribution is normalised such that it sums to unity on every iteration of the gradient descent loop.
When fitting undithered data with a large number of gradient descent iterations, these methods tend to concentrate flux sparsely in a few bright pixels. To mitigate this, the source distribution is convolved with a small Gaussian kernel ($\sigma\approx0.25\,\text{pix}$) each fitting epoch.
The image parameters (pixel values) are fit in log-space to enforce image positivity. After the fit, the log distribution is simply exponentiated to provide the final reconstructed image~$\hat{I}$.

The primary feature that differentiates the methodologies introduced here is the data product used in the calculation of the likelihood. That is, which data product is $\mathbf{y}$ in \Cref{eq:bayes2} and \Cref{eq:bayes1}.

\subsection{Method 1: Image Plane Method}
\label{sec:method1}
\begin{figure}
\begin{center}
\includegraphics[width=1.\textwidth]{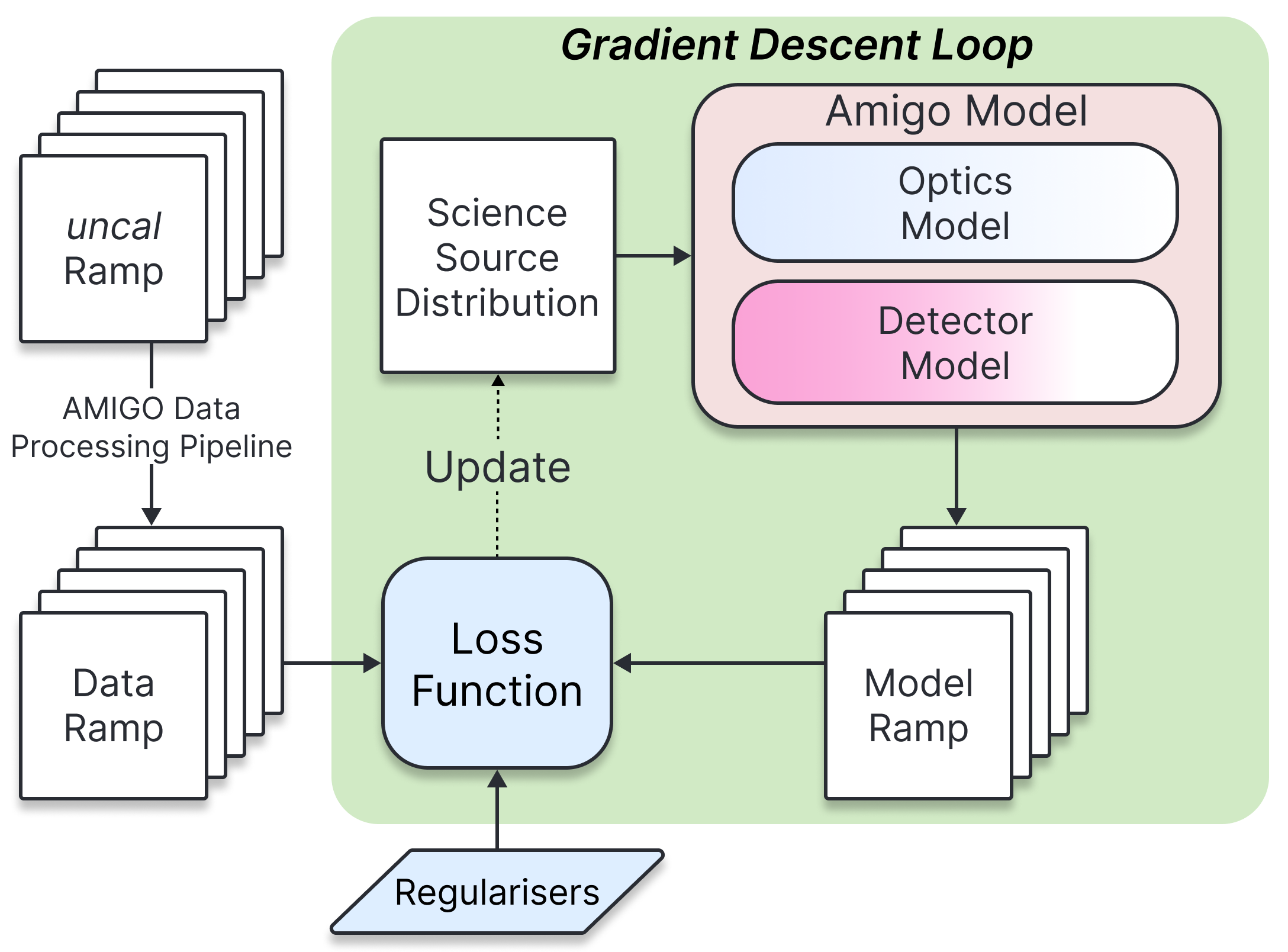}
\end{center}
\caption 
{ \label{fig:method1}
Flow diagram depicting the Method 1 image plane-based reconstruction process. The source is modelled via a convolution with a source distribution array and is fit to the ramp data with gradient descent. The final image is obtained from the science source distribution after a specified number of gradient descent iterations are complete.} 
\end{figure}

In the image plane method, the data product $\mathbf{y}$ that is used for the likelihood is the ramp data. 
The interferogram is modelled by a convolution between the source distribution and the \ac{psf}, as opposed to complex visibilities which is the \amigo default. Complex visibilities are not used at all in this method, thus the fit is performed entirely in the image plane. This fitting process is depicted in \Cref{fig:method1}. The parameters of the \amigo model which are fit to the ramp data are the
\begin{itemize}
\item positions (per exposure),
\item fluxes (per exposure),
\item spectra (per target per filter),
\item log distribution $\log_{10}\mathcal{I}$ (fit only to the science target per filter, in log space to enforce image positivity),
\item mirror aberrations $\mathbf{Z}$ (fit only to the calibrator target per filter).
\end{itemize}

By avoiding the Fourier domain, Method 1 targets the scenario in which the limiting factor affecting image quality is detector non-linearities. Additionally, unlike closure phase methods, this image plane method inherently incorporates low spatial frequency information through short redundant baselines, as discussed in the following section.

\subsection{Method 2: DISCO Method}
\label{sec:method2}

\begin{figure*}[ht]
\centering
\includegraphics[width=1.\textwidth]{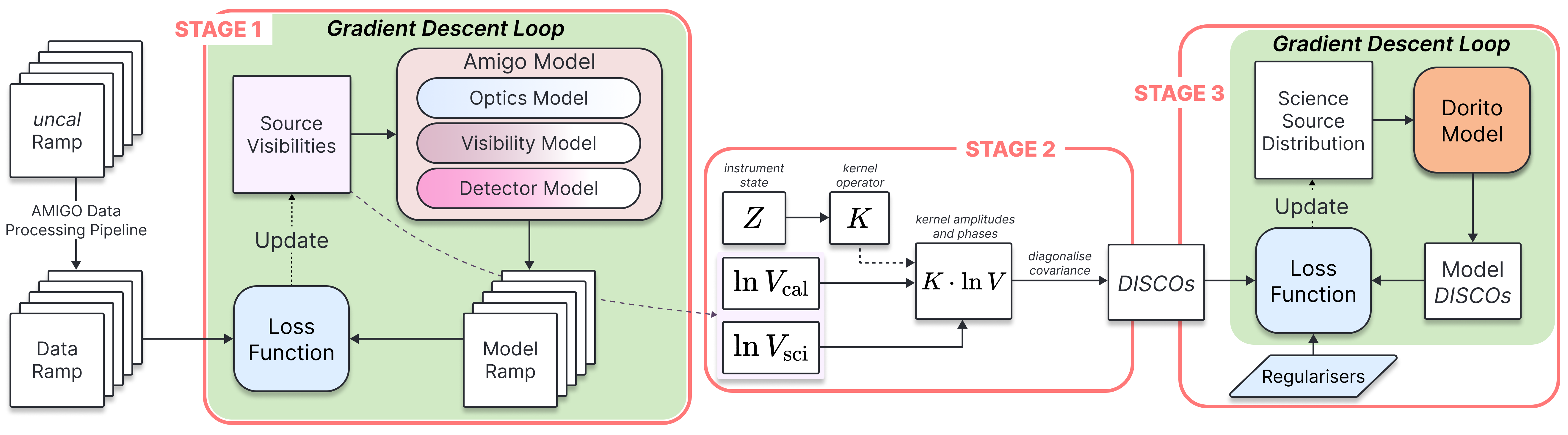}
\caption{Flow diagram depicting the Method 2 DISCO-based image reconstruction process, broken into three stages. In the first, the source distribution is forward-modelled with complex visibilities and then transformed to the DISCO basis in the second. This is identical to the fitting processes in \citet{Desdoigts2025}, and is described in further detail there. Thirdly, the image reconstruction takes place as a source distribution array is fit to the reduced DISCOs.}
\label{fig:method2}
\end{figure*}

As opposed to Method 1, the \ac{disco} method uses the Fourier domain to characterise the source distribution. While using Fourier quantities is a more typical approach to image reconstruction, the use of \acsp{disco} is a novel method introduced by \amigo.

While \ac{ami} generates closure phases \citep{closure_phase} between triplets of its holes, the holes themselves have a substantial width in comparison to their separation.
Despite the arrangement of the \ac{nrm} holes into a non-redundant configuration, the non-zero finite hole area results in short redundant baselines within an individual hole which transmit low spatial frequency modes.
We would like to access this information to accurately recover the low spatial frequencies in the image, however closure phases are only valid for non-redundant apertures.
Furthermore, one of the findings of \citet{Desdoigts2025} is that the aperture mask is not quite conjugate to the telescope pupil plane and a small Fresnel defocus is required; this near-field effect means that the closure phases are no longer invariant. As a result of this redundancy and defocus, the standard closure phases are no longer robust against wavefront error, calling for a more general treatment. 

We therefore adapt the kernel phase and amplitude formalism \citep{kernel_phase,Pope_2013,kernel_amp}, which generalises closure invariants to redundant apertures:

\begin{equation}
\Phi^{\text{meas}} = \Phi^{\text{sky}} + \mathbf{J} \cdot \phi,
\end{equation}

\noindent with a Jacobian matrix $\mathbf{J}$ that describes the transfer of phase errors $\phi$ on the pupil into $u$-$v$ visibility phase measurements $\Phi^{\text{meas}}$ of the astrophysical information. 
That is, $\mathbf{J}$ quantifies the response of the $u$-$v$ plane to perturbations of pupil plane phase error modes; here we assume these phase errors can be described with a linear combination of $n$ Zernike modes. Thus $\mathbf{J}$ is the matrix for partial derivatives of basis vectors spanning the $u$-$v$ plane with respect to the first $n$ Zernike polynomials.

If we then generate a null operator $\mathbf{K}$ (also known as kernel operator) for the Jacobian matrix $\mathbf{J}$, we can use this to generate self-calibrating kernel phases

\begin{equation}
\mathbf{K}\cdot\Phi^{\text{meas}} = \mathbf{K}\cdot\Phi^{\text{sky}} + \underbrace{\mathbf{K}\cdot\mathbf{J} \cdot \phi}_{=0}\\,
\end{equation}

\noindent i.e. the null operator nulls out the Zernike modes that transfer phase errors $\phi$ to the $u$-$v$ plane. 
This leaves the kernel phases $\mathbf{K}\cdot\Phi^{\text{meas}}$ which contain less information than $\Phi^{\text{meas}}$ itself as all phases that could be expressed as a linear combination of the first $n$ Zernike polynomials are mapped to zero. However this is an advantageous trade-off as the kernel phases should be invariant to wavefront phase error.
Although we describe the phase as an expansion in the Zernike basis, the mathematics is the same as the classic kernel phase derivation of low amplitude piston phases on small sub-apertures. 
We do not treat amplitude error in the pupil; in developing \amigo we allowed pupil transmission to be a free parameter, but did not consistently recover significantly nonzero values.
Furthermore, by including a wavelength term in the Jacobian we can also null the effects of spectral miscalibration when applying the kernel operator.


We can generate the transfer matrix $\mathbf{J}$ not from a geometric model and \ac{fft}, which is prone to sampling and interpolation issues \citep{Martinache2020}, but rather by forward-modelling the system and using \texttt{Jax} \ac{autodiff} to generate the Jacobian \citep[following][]{kernel_phase_ad}.

In the \ac{disco} method introduced in \citet{Desdoigts2025}, the \amigo forward-model with Fresnel diffraction means that phase errors in the pupil generate both phase and amplitude variations in the visibilities, and therefore kernel phase and kernel phase-to-visibility matrices. We extract complex visibilities through forward-modelling with Bayesian inference, by optimisation and then applying the Laplace approximation to estimate the covariance of these parameters \citep{Kass1991}. We keep track of covariance as we generate these kernel observables, and then re-diagonalise so that they are statistically independent following \citet{ireland_2013}. We refer to these observables as \acp{disco} to reflect that these are neither the previously-used kernel phase nor amplitude observables, as they include not just near-field effects but also spectral miscalibration. 

To perform image reconstruction from \acp{disco}, there are three separate stages, with two separate fitting loops. This method is depicted in \Cref{fig:method2}. The first stage is a ramp fit similar to Method 1. However, rather than modelling a source distribution directly, the resolved component is modelled with complex visibilities. Thus the \amigo model parameters that are fit to the ramp in the first stage are the

\begin{itemize}
\item positions (per exposure),
\item fluxes (per exposure),
\item spectra (per target per filter),
\item log visibility amplitudes $\operatorname{Re}\left(\ln V\right)$ (per exposure),
\item visibility phases $\operatorname{Im}\left(\ln V\right)$ (per exposure),
\item mirror aberrations $\mathbf{Z}$ (fit only to the calibrator star).
\end{itemize}

The second stage is the transforming of the complex visibilities from the $u,v$ basis to the \ac{disco} basis. This involves applying the kernel operator $\mathbf{K}$ (which is generated from the fitted mirror aberrations $\mathbf{Z}$), then diagonalising by the tracked covariance matrix. The third and final stage is where the image reconstruction is performed. Rather than propagating a source distribution through the AMIGO model as in Method 1, the source distribution is transformed to the \ac{disco} basis and fit with the reduced \acp{disco} from the second stage.

\acp{disco} are designed to be resistant to wavefront error, and thus Method 2 is expected to be most effective when limited by residual wavefront error. Since \acp{disco} are invariant to Zernike tip/tilt modes by design, they are also invariant to translations in the image plane. This introduces a heavy covariance to the pixel parameters in the third stage fit, as there is nothing to constrain the image centroid position. To assist convergence, a tight prior is added enforcing the image centroid position to be at the image centre.

\section{Datasets}\label{sec:datasets}
In this work, the two methods described in \Cref{sec:methods} are each applied to \ami datasets of three different science targets. The targets were chosen as they are publicly accessible datasets that offer a diversity of astrophysical morphologies to test our methods. NGC~1068 has an extended complex structure behind a bright central core, Io is heavily resolved with large regions of uniform flux and scattered point sources, whereas WR~137 pushes the angular resolution limits of our methods as it features a bright linear plume that is only partially resolved.
The interferograms for each science target and its associated PSF reference star are displayed in \Cref{fig:data}. All observations were taken on NIRISS' SUB80 sub-array with the NISRAPID readout pattern. All targets were observed through some subset of the following three near-infrared filters:
\begin{itemize}
\item \fblue \textcolor{color1}{($\lambda=3.825$\,\si{\micro\meter}, $\Delta \lambda=0.205$\,\si{\micro\meter})},
\item \fgreen \textcolor{color3}{($\lambda=4.285$\,\si{\micro\meter}, $\Delta \lambda=0.213$\,\si{\micro\meter})},
\item \fred \textcolor{color2}{($\lambda=4.815$\,\si{\micro\meter}, $\Delta \lambda=0.303$\,\si{\micro\meter})}.
\end{itemize}

The other filter available on AMI, \textcolor[HTML]{7B61FF}{F277W}, is not Nyquist-sampled, and has not yet been used for imaging.
Additionally, the neural network predicting \ac{bfe} charge bleeding in the \amigo base model was trained on calibration data which only reached $\sim45\%$ full pixel well depth (or $\sim30\,\rm{k}$ counts). Hence it is not reliable in predicting \ac{bfe} above this, so here we truncate the ramp of the datasets to stay below this limit by discarding the high flux groups.

\subsection{NGC~1068}
NGC~1068, or M77, is the prototypical Seyfert 2 \ac{agn} located at $\sim14\,\rm{Mpc}$ distance, powered by accretion onto a \ac{smbh} of $\sim 3 \times 10^{7}\,M_{\odot}$ \citep{Ford2014}. Seyfert 2 \ac{agn} are believed to be observed edge-on, obscuring our view of the central engine at short wavelengths but yielding an IR-bright source in the dusty parsec-scale torus \citep{Padovani17}. The details of how \ac{agn} such as NGC~1068 are fuelled remains unclear as does the effect of luminous feedback from an \ac{agn} on its host galaxy.

NGC~1068 is also confidently associated with the detection of TeV neutrinos, implying the interaction between hadrons in a relativistic outflow and surrounding dense material at luminosity $L_{\nu} \sim 3 \times 10^{42} {\rm \,erg/s}$ \citep{Inoue20,IceCube22}, potentially from accreting stellar mass black holes embedded within the \ac{agn} disk \citep{Tagawa23}.

NGC~1068 and its calibrator HD~17130 were observed on two separate occasions (Sept 2024, Aug 2025) by \ami as part of the GTO program 1260 (PI: K.S. Ford). Both targets were observed with all three filters (\fblue, \fgreen, and \fred) in a five-point subpixel dither pattern. 
The first epoch of science data saturated the detector with 21, 25, 20 groups and 141, 112, 91 integrations for the \fblue, \fgreen and \fred filters respectfully. The second epoch used 7 groups for all filters and 134, 139, 100 integrations; for both epochs, we only fit to the first 4 groups of the ramp to stay below $\sim30\,\rm{k}$ counts.

The region of interest to these observations has been imaged by \ac{lbti} in the mid-IR, which resolved a general N-S outflow, with clumps more evident in the N outflows \citep{Isbell25}; we shall directly compare our image reconstruction results with that of \citet{Isbell25}.

\subsection{Io}
Io is the innermost of the Galilean moons of Jupiter and is hailed as the most volcanic place in the solar system. This is a result of the extreme tidal forces from Jupiter and gravitational interactions with the other Galilean moons as they exhibit orbital resonance. Erupting volcanoes are always visible on the surface of Io, shining especially bright in infrared light. The infrared surface of Io has been well characterised by the Juno spacecraft, which has been orbiting Jupiter since 2016 \citep{adriani2017jiram, davies_junoexam}.

On August 1st 2022, Io and its calibrator HD~2236 were observed using the \ami mode. This was part of a suite of observations of the Jovian system across a broad range of observing modes to demonstrate \ac{jwst}'s capability for solar system science (DD-ERS program 1373, PI: I. de Pater). Observations of Io were taken only through the \fgreen filter in five subpixel dither positions, each consisting of 45 groups and 100 integrations. 
This also saturated the detector, so only the first 18 groups of the ramp were used in order to stay below $\sim30\,\rm{k}$ counts.

The \ac{jwst}-Io-Sun phase angle is always small (10.3\textdegree, for these data) due to \ac{jwst}'s solar proximity relative to the Jovian system, so we do not expect to resolve a terminator.
Each observation was taken $8.55\,\si{\minute}$ apart on average, during which Io rotates $1.21\degree$ on its axis \citep{king_2023_planetmapper}. 
This temporal rotation is significant enough to be resolved between the dithered exposures. Thus instead of jointly fitting to all five dithered exposures, we treat each exposure as its own undithered image to create a time-evolving image.
The calibrator star HD~2236 was observed with the identical observing configuration, at 4 primary dither positions each with 12 groups and 8 integrations, amounting to only $32.04\,\si{\second}$ total exposure time.

In a recent analysis of these data, \citet{Sanchez-Bermudez2025} applied several neural network and filter architectures to image reconstruction, detecting several hot-spots and identifying these with volcanic features on Io. They do not cleanly resolve the limb of the moon against the black backdrop of space, and registration of the reconstructed image to the coordinate system of Io is made with respect to these bright features. One volcanic feature designated ``Seth Patera'' was found to be extended, and the overall surface was seen to rotate over the course of the exposure.

\subsection{Wolf-Rayet 137}
The colliding wind binary system WR~137 is comprised of a carbon-rich (WC-type) Wolf-Rayet together with a hot O9-spectrum companion \citep{WR1371985,WR1372001}. Classified as an episodic dust producer, variation in the strength of the colliding wind interaction mediated by orbital phase results in a $\sim13$\,yr periodicity evident in the long term infrared light curve.
In common with other such ``pinwheel'' systems, the dust streams outwards carried on the fast winds, creating expanding structures that engrave the physics of wind and orbit in nested shells with spiral morphology that may be bright in the near- and mid-infrared.

\input{WR137SCHEDULE}

Over three observing runs in 2022, WR~137 and its calibrator target HD~228337 were observed by \ami as part of DD-ERS program 1349 (PI: R. Lau). Primary-dithered observations were taken through the \fblue and \fred filters. These observations did not significantly exceed $30\rm{k}$ counts so all groups up the ramp were used here. Further details about the observation configuration/schedule are given in \Cref{tab:wr137_schedule}.

This schedule is significantly more complicated due to a mirror tilt event during the first observing run, after almost all the science data had been taken but none of the calibrator data. After the mirror tilt event, this run was abandoned and rescheduled without taking any calibrator observations. Such a rapid change in optical state renders PSF reference data unhelpful: it would embody a differently aberrated wavefront to the science target. The second observing run was performed two days after the mirror tilt event, now including the calibrator and science target.

These observations have previously been analysed by \citet{2024ApJ_RyanLau}, using a variety of extraction pipelines \citep[\texttt{ImPlaneIA}, \texttt{AMICAL}, and \texttt{SAMPip}:][]{ImPlaneIA,amical,sampip} and image reconstruction algorithms \citep[\texttt{BSMEM}, \texttt{Squeeze}, and \texttt{IRBis}:][]{BSMEM,squeeze,irbis}.

\section{Results}

Reconstructions of all three targets are shown in the final row of \Cref{fig:data}. For all reconstructions, the source distribution was initialised as a uniform array so that the reconstruction process remains unbiased. The images of NGC~1068 and Io were reconstructed using Method 1 with TV regularisation, while WR~137 was reconstructed using Method 2 with MaxEnt regularisation. 

\subsection{NGC~1068 Results}

\begin{figure}[h]
\begin{center}
\includegraphics[width=1.\textwidth]{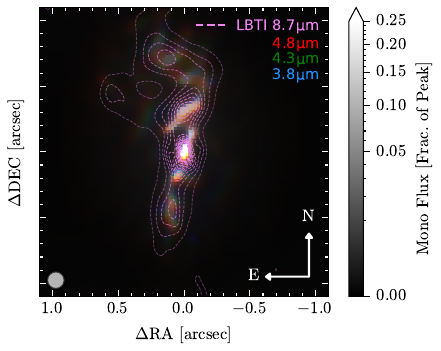}
\end{center}
\caption 
{A comparison of the NGC~1068 images recovered in \ami with the \ac{lbti} observations convolved by \citet{Isbell25}, updated with recent data by private communication. The three bands of \ami data in \fred, \fgreen, and \fblue are represented as red, green, and blue in a false-colour image, with the \ac{lbti} 8.7\,$\mu$m image flux overlaid as contours. The beam size circle in the lower left indicates $\lambda/2D$. The bright parts of the \ami colour image are close to white, indicating a consistent recovery across all three bands, and they track the bright parts of the \ac{lbti} image very closely, indicating both sets of data are independently recovering the same structure.} 
\label{fig:lbti}
\end{figure}

Fig.~\ref{fig:lbti} shows our reconstructed image with \ac{lbti} results from \citet{Isbell25} overlaid in contours. The \ac{lbti} image reconstruction at $8.7\,\micro$m neatly covers our entire reconstructed image. We find excellent agreement overall, with some additional details apparent (and noting that exact agreement is not expected due to the significantly different observing wavelength). The obscuring torus at the heart of NGC~1068 lies $\sim -0.1^{"}$ below the bright central core of this image in a SE-NW orientation \citep{Gamez2022}. The right-hand boundary of the \ac{lbti} image, coincides approximately with one edge of the $\sim 60^{\circ}$ opening angle [$O_{\rm{III}}$] emission cone observed by \citet{Macchetto94}, which presumably defines the region of low density gas exposed to the \ac{agn} radiation field. The 5\,GHz radio emission from the $~\sim 100\,{\rm pc}$ nuclear radio jet extends from the torus location to the NNE in this image (through $0.2^{"},0.6^{"}$). Thus, the central brightest part of our image is an off-axis hotspot around $\sim 10\,{\rm pc}$ projected distance from \ac{smbh} due to outflow interaction with the dusty environment outside the torus. 

Dusty gas is likely the source of much of the reconstructed emission signatures. From \citet{MorNetzer12}, we might expect the dust sublimation radius $R_{\rm sub}$ to lie a distance
\begin{equation}
R_{\rm sub} \sim 0.4{\rm pc} \left(\frac{L_{\rm bol}}{10^{45}{\rm erg/s}} \right)^{1/2} \left(\frac{T_{\rm sub}}{1500{\rm K}}\right)^{2.6}
\end{equation}
from the \ac{smbh}. Thus, dust at $\sim 10\,{\rm pc}$ should survive being exposed to the inferred radiation field and reprocess radiation into the IR. Part of the hotspot emission may of course include both an overdensity of dusty gas as well as shock-heating and destruction of any dust overdensity. 
Since outflowing winds are typically $\mathcal{O}(10^{3})\,{\rm km/s}$, a hotspot at $\mathcal{O}(10)\,{\rm pc}$ implies that this episode of the \ac{agn} has been running for at least $10^{4}\,{\rm yr}$. Likewise the overdensities within $\sim 100\,{\rm pc}$ illuminated by outflow imply that this \ac{agn} episode has been running for $\sim0.1\,{\rm Myr}$. Assuming a uniform accretion rate during that time implies a mass accretion of $\sim 10^{4}\,M_{\odot}$ onto the \ac{smbh}, with a sizeable (comparable) mass of gas flung back outward in an outflow. More generally, drawing all of these observations together suggests that the feedback from this $L_{\rm bol} \sim 10^{45}\,{\rm erg/s}$ \ac{agn} is not powerfully impacting the surrounding galaxy. Dusty gas is still present within, or at the boundary of, the $\sim 60^{\circ}$ outflow region from the \ac{agn}.



\subsection{Io Results}

\begin{figure*}[h!]
\begin{center}
\includegraphics[width=1\textwidth]{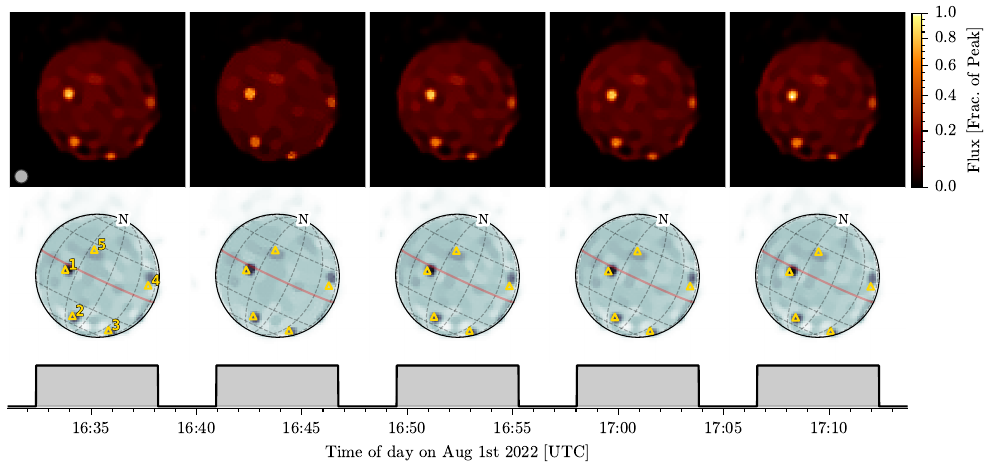}
\end{center}
\caption{
Reconstructions of each of the five exposures of Io, chronologically left to right. 
\textit{Top row}: shows the reconstructed images, which were reconstructed with Method 1 using TV regularisation. The rotation of Io on it's axis becomes visible when displayed as a time series animation, \href{https://maxecharles.github.io/dorito/}{hosted online}. 
\textit{Second row}: shows those same images with an ephemeris overlay, including the expected positions of five Ionian volcanic surface features. These are 1) Seth Patera, 2) P197, 3) Masubi, 4) Leizi Fluctus, 5) Amirani \citep{Davies2024, VolcanismOnIo}. 
\textit{Bottom axis}: a time axis showing the times over which the exposures were taken.
} 
\label{fig:ios}
\end{figure*}

We have reconstructed images of Io using Method 1 with TV regularisation, with all five epochs shown in the top row of \Cref{fig:ios}. Io is \textit{heavily} resolved as seen by \ami, and visibility extraction in the forward-modelling approach has performed poorly; we have been unable to get any adequately converged image using Method 2 with a variety of initialisations and hyperparameter tunings. In the Method 1 images we detect five known Ionian volcanic surface features at their expected positions: 1) Seth Patera, 2) P197, 3) Masubi, 4) Leizi Fluctus, 5) Amirani \citep{Davies2024, VolcanismOnIo}. There is a visual agreement between the expected positions and reconstructed bright points on Io's surface, as was also found by \citet{Sanchez-Bermudez2025}. 

Each volcano appears to be surrounded by a valley of flux. These are presumably not physical and are simply ringing artefacts due to minor remaining \ac{psf} miscalibration. This effect is somewhat suppressed by the TV regularisation; this can be seen in \Cref{fig:io-Lcurve} as the less regularised images exhibit more severe ringing artefacts.

Over time, the moon can be seen to rotate subtly on its axis --- a feature for which the model had no a priori information or bias. This can be seen by tracking the Ionian meridians in the ephemeris overlay, but it is more identifiable when the exposures are played as frames in a time series. The detection of rotation is discussed in detail by \citet{Sanchez-Bermudez2025}, and we do not seek to interpret this scientifically; it is unambiguously consistent with the rotating frame of Io visually in Figure~\ref{fig:ios} and in an animation of the rotation \href{https://github.com/maxecharles/dorito}{hosted online}.

We do not see evidence for a more diffuse emission around the brightest feature Seth Patera, which is observed both in the \citet{Sanchez-Bermudez2025} \ami reduction and corresponding Keck observations \citep{Marchis2002,deKleer2019}. In the Keck imaging presented therein, the emission is more extended at shorter wavelengths, and may be less apparent in \fgreen. It is possible that the published extended structure around this hotspot may be either miscalibrated \ac{bfe} or \ac{psf} structure.



\subsection{WR~137 Results}
We have reconstructed images of WR~137 using Method 2 with MaxEnt regularisation, shown in Figure~\ref{fig:data}. This image was jointly fit exclusively to exposures taken after the mirror tilt event on the second and third observing epochs (Jul 15th and Aug 8th, see \Cref{tab:wr137_schedule}).

\begin{figure}[h]
\begin{center}
\includegraphics[width=1.\textwidth]{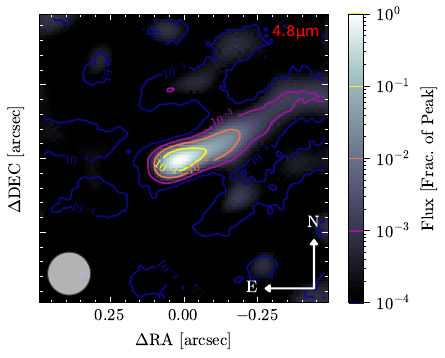}
\end{center}
\caption 
{Comparison between a reconstructed image of WR~137 from \citet{2024ApJ_RyanLau} (contours) and this work (image), shown on a logarithmic colour stretch. Both images are reconstructed from the same dataset using the \fred exposures after the mirror tilt event. The image from \citet{2024ApJ_RyanLau} was reconstructed using \texttt{Squeeze} on observables extracted with \texttt{AMICAL}, which we considered the cleanest of the reconstructions given by \citeauthor{2024ApJ_RyanLau}. It was convolved with a small Gaussian filter ($\sigma=7.42\,\si{\milli\arcsecond}$) to declutter the contour lines for visualisation purposes. The image from this work was reconstructed with \acp{disco} using Method 2. The beam size circle in the lower left indicates $\lambda/2D$. The two images closely match in structure, including very faint patches of flux near $\sim10^{-4}$ which are likely common non-physical miscalibration artefacts.}  
\label{fig:lau}
\end{figure}

As shown by the contour overlay comparison in \Cref{fig:lau}, our results independently confirm the earlier imagery of \citet{2024ApJ_RyanLau}. Our reconstruction shows a linear plume along an ESE-WNW orientation, present in all \citet{2024ApJ_RyanLau} reconstructions. The length and orientation of this plume are consistent with models for a colliding-wind binary pinwheel nebula seen side-on. For further astrophysical interpretation, the reader is referred to the excellent discussion found in \citet{2024ApJ_RyanLau}.

\begin{figure*}
\centering
\includegraphics[width=.9\textwidth]{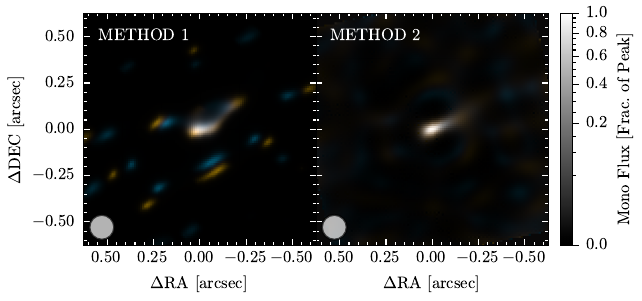}
\caption 
{Images of the WR~137 dataset affected by a tilt event, deconvolved with different methods. \textit{Left}: deconvolved with Method 1, in the pixel domain. The tilt event and consequent \ac{psf} misspecification mean that there are chromatic speckles systematically introduced into the image. \textit{Right}: deconvolved with Method 2, from \acp{disco}. Even though this method has performed poorly on the other datasets, it accurately retrieves the simple plume shape of WR~137 without \ac{psf} artefacts, as the \acp{disco} act like closure phases and are insensitive to changes in the optical path difference between target and calibrator by design.} 
\label{fig:MTE}
\end{figure*}

Additionally, to assess the robustness of \acp{disco} against wavefront error, we reconstructed images using both Method 1 and 2 from the science exposures taken on the first observing epoch (Jul 13) before the mirror tilt event. As there was no calibrator target observed that day, these science exposures were calibrated by the calibrator exposures taken \textit{after} the mirror tilt event on Jul 15. Both reconstructed images are shown in \Cref{fig:MTE}. 

As expected, Method~2 proved to be superior in dealing with wavefront error. The image reconstructed from Method~1 is plagued by numerous spectrally-dependent non-physical miscalibration artefacts. By contrast, the Method~2 image is relatively free from these artefacts and is visually similar to the image created from the other two epochs in \Cref{fig:data}.

\section{Conclusions}

This paper has presented \dorito, an image reconstruction pipeline for recovering complex scenes from \ac{niriss}/\ami observations. It has demonstrated recovery of diffraction-limited images, which are consistent with the highest-quality ground-based and space-based images of the same targets and of a generally improved quality compared to published results from the same data. 

As the mask holes are relatively large given their separations, \ami passes a large number of Fourier modes to the interferogram and the image reconstruction problem is not nearly so underconstrained when working with forward-modelling the full $u$-$v$ plane as it is when only working with the 21 baselines. In general, we have been able to deconvolve acceptable images of the moderate-contrast sources presented in this paper without regularisation at all, and have only noted modest benefits from adding these terms in. We expect regularisation to become more important at high contrast (where the signal-to-noise ratio is lower) and for very complex scenes with significant high spatial frequency detail (where it is therefore necessary to incorporate more Fourier modes).

As noted by \citet{DeFurio2025}, the frequent wavefront sensing measurements taken by \ac{jwst} make it possible --- in principle --- to do calibrator-free interferometry. With a drift rate of $<10$\,nm rms per 48\,h \citep{Lajoie2023}, aside from tilt events it is likely that our gnosis of the wavefront at any point in time is almost as good as would be achieved with a point source reference in the absence of any other information. While for the time being we recommend CAL-SCI or CAL-SCI-CAL observing sequences in line with the \ac{jwst} proposal requirements, we recommend benchmarking this option as a step towards significantly reducing the time overheads and constraints required for \ami observations.

This work has revealed that it is not in general necessary to use \acp{disco}; rather the more straightforward approach of modelling the source in the image plane performs well in most cases. Nulling all the linear combinations of visibilities affected by wavefront miscalibration, as performed by the \acp{disco} algorithm, is extreme. While in the \amigo companion paper \citep{Desdoigts2025} we find this enhances our sensitivity to faint point sources, we believe that in the present paper we lose too many Fourier modes for this to be a viable approach for imaging complex scenes. Nevertheless if we appropriately propagate wavefront error from a calibrator or from routine wavefront sensing operations, we may be able to avoid fully nulling these modes while retaining benefits of self-calibration. 

Our ongoing recommendations for \ami proposals agree with those set out in the companion paper: maximise the number of groups subject to keeping the maximum well depth below $\sim30\,\rm{k}$ counts. Observe calibrators of a similar brightness to the science target ensuring integration to a similar well depth. Similar recommendations were made by \citet{2024_Sallum}. We disagree with the practice in many \ami observations to date of ensuring the science target and calibrator are placed on the same subpixel position and kept as still as possible; to the contrary we strongly recommend using as many subpixel dithers as is practical, to better separate detector effects from astrophysical information. 


There is a wealth of archival observations that will benefit from \amigo, \dorito, or similar approaches. There are many cases where it is important to observe faint point sources at separations close to the diffraction limit, but this work shows that we can also recover complex scenes effectively. In GTO~1242, we hope to be able to refine the fits to PDS~70 published in \citet{doripds70}, and extend this to the other transition disks HD~135344~B \citep{Cazzoletti2018} and HD~100546 \citep{Quanz2015} observed in the same program. Beyond archival data, there are a wealth of protoplanetary disks to observe with similarly impactful science \citep[e.g. in the DSHARP survey:][]{dsharp}. The example of NGC~1068 shows that it will be valuable to observe \ac{agn} inner environments, and quasar host galaxies such as 3C~273 \citep[following][]{Ren2024}. While it is possible to monitor the volcanism on Io with Keck adaptive optics imaging, \ami may prove to be a valuable complement.


\ac{rml} imaging has long been a standard approach for image reconstruction in astronomy, and it has been shown to perform well in the present context. Nevertheless, it is far from the most modern or sophisticated approach to image deconvolution, and while a pixel basis performs well here, there is no reason to think this is optimal. Future work will include investigation of different bases, including wavelet bases which are currently being implemented in \dorito. Machine learning also offers many paths to improvement of \dorito. The literature here is wide-ranging, but we expect there will be great value in approaches based on score-based priors and likelihoods \citep[e.g.][]{Adam2025,Dia2025}, deep image priors \citep[e.g.][]{Feng2024}, information field theory \citep{Ensslin2019}, autoencoders \citep[notably the \ami Io work by][]{Sanchez-Bermudez2025}. This is likely to be especially impactful in resolving sources very close to the inner working angle of \ami and in generalising this approach to differentiable rendering of coronagraphic data \citep[as in][]{Feng2025}, where the problem is much more nonlinear and there is much greater distortion of the scene close to the inner working angle.

\section{Code and Data Availability}
\label{sec:data}

All codes and data used to produce this work are publicly available and open source. The \amigo model and pipeline are hosted \href{https://github.com/LouisDesdoigts/amigo}{on GitHub}. The \dorito model is hosted \href{https://github.com/maxecharles/dorito}{on GitHub}, along with a series of \href{https://github.com/maxecharles/dorito_notebooks}{example notebooks} used in this paper's results. All data are available from \ac{mast} under the appropriate proposal numbers; in this pipeline paper, we have relied on publicly available data with no proprietary data included, with the exception of the latter observing epoch of NGC~1068 at the time of this submission. We encourage researchers to adapt and apply \amigo and \dorito to their own interferometric datasets, and to contact the development team with questions and contributions.

\begin{acknowledgement}
The authors thank Imke de Pater and Joel S\'anchez Berm\'udez for their valuable assistance in identifying Ionian surface features and John Stansberry for his work on Io observations; Jacob Isbell for sharing the latest LBTI reconstructions of NGC 1068; and Ryan White for the consultations about colliding wind-binary morphologies.

BP, PT, and SR have been supported by the Australian Research Council grant DP230101439 and BP by DE210101639; and LD and MC have been supported by the Australian Government Research Training Program (RTP) award. We are grateful to the Australian public for enabling this science. BP and SR would like to thank the Big~Questions~Institute for their philanthropic support. Development of \dlux has been supported by the Breakthrough Foundation through their Toliman project as a part of the Breakthrough Watch initiative. DJ is supported by NRC Canada, and DB and DJ acknowledge the support of the Natural Sciences and Engineering Research Council of Canada (NSERC). BM \& KESF are supported by NSF AST-1831415, NSF AST-2206096 and Simons Foundation Grant 533845.

We acknowledge and pay respect to the traditional owners of the land on which the University of Sydney, University of Queensland and Macquarie University are situated, upon whose unceded, sovereign, ancestral lands we work. We pay respects to their Ancestors and descendants, who continue cultural and spiritual connections to Country.  
We acknowledge and respect the L\textschwa\'{k}\textsuperscript{w}\textschwa\textipa{\ng}\textschwa n (Songhees and Esquimalt) Peoples on whose territory the University of Victoria stands, and the L\textschwa\'{k}\textsuperscript{w}\textschwa\textipa{\ng}\textschwa n and \underline{W}SÁNEĆ Peoples whose historical relationships with the land continue to this day.


This work is based on observations made with the NASA/ESA/CSA \textit{James Webb Space Telescope}. The NIRISS instrument was funded
by the Canadian Space Agency and built in
Canada by Honeywell. The \amigo base model is constructed from observations
associated with JWST program CAL~4481 (PI: A. Sivaramakrishnan). The data were obtained from \ac{mast} at \ac{stsci}, which is operated by the Association of Universities for Research in Astronomy, Inc., under NASA contract NAS 5-03127 for JWST. 

This work was performed on the OzSTAR national facility at Swinburne University of Technology. The OzSTAR program receives funding in part from the Astronomy National Collaborative Research Infrastructure Strategy (NCRIS) allocation provided by the Australian Government, and from the Victorian Higher Education State Investment Fund (VHESIF) provided by the Victorian Government.

This research made use of \dlux \citep{dLuxI,dLuxII}; \texttt{Jax} \citep{jax}; \texttt{equinox} \citep{equinox}; \optax  \citep{optax2020github}; \texttt{planetmapper} \citep{king_2023_planetmapper}; and \texttt{matplotlib} \citep{matplotlib}.

\end{acknowledgement}


\bibliography{report}

\appendix

\section{Other L-curves}
\label{appendix:lcurves}

\begin{figure*}[h!]
\begin{center}
\includegraphics[width=\textwidth]{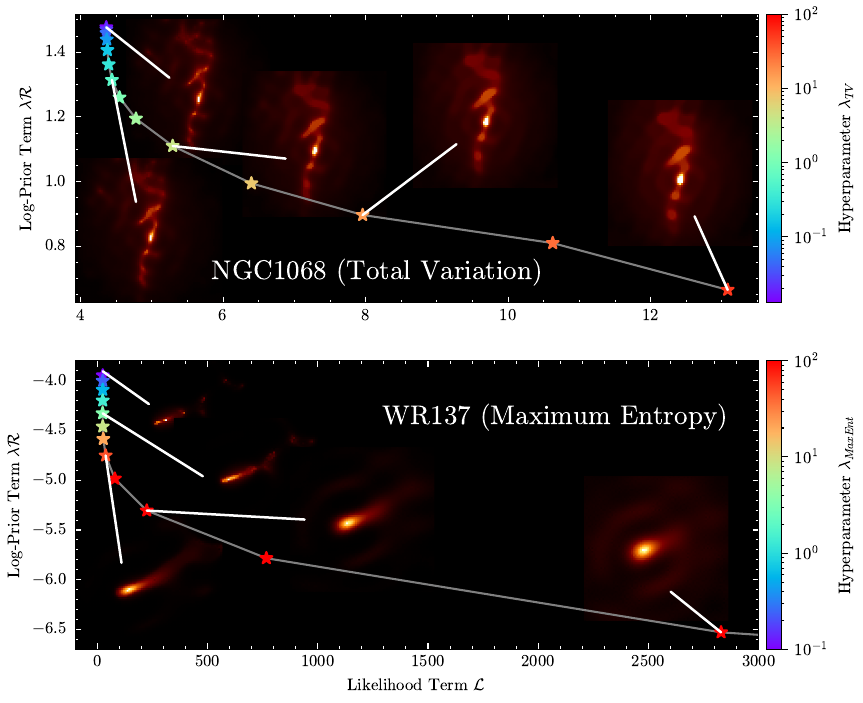}
\end{center}
\caption{L-curve diagram used to select the optimal regularisation parameters for NGC~1068 (top) and WR~137 (bottom). Each point on the curve is the balance between regulariser term $\mathcal{R}$ and likelihood term $\mathcal{L}$ of a converged image reconstruction for a different regularisation hyperparameter value. NGC~1068 is regularised with total variation, and WR~137 is regularised with maximum entropy. Shown are several reconstructed images corresponding to different points along the curve.} 
\label{fig:other-Lcurve}
\end{figure*}

L-curve plots for sources NGC~1068 and WR~137 are shown in \Cref{fig:other-Lcurve}, with the corresponding curve for Io shown in \Cref{fig:io-Lcurve}. Unregularised images of NGC~1068 largely resemble those at the elbow of the L-curve. Compared to Io, there are significantly less non-physical ringing artefacts needing to be suppressed with regularisation. This may imply the NGC~1068 observations were better calibrated than the Io observations, which is unsurprising given the higher data quality of the NGC~1068 calibrator exposures. Increasing TV regularisation resulted in smoother images as expected but did not largely change the morphology of the \ac{agn}.
Contrastingly, regularisation appears to be necessary for WR~137. With no regularisation WR~137 collapses into a few loosely connected points of flux which would be impossible to resolve, as a result of overfitting.

\end{document}

%% file: acronyms.tex
\DeclareAcronym{jwst}{
  short = JWST,
  long = \textit{James Webb Space Telescope}
}

\DeclareAcronym{agn}{
  short = AGN,
  long = Active Galactic Nucleus,
  cite = {Antonucci1985,Jaffe2004}
}

\DeclareAcronym{fft}{
  short = FFT,
  long = Fast Fourier Transform
}

\DeclareAcronym{ami}{
  short = AMI,
  long = Aperture Masking Interferometer
}

\DeclareAcronym{nrm}{
  short = NRM,
  long = Non-Redundant Mask
}

\DeclareAcronym{psf}{
  short = PSF,
  long = point spread function
}

\DeclareAcronym{niriss}{
  short = NIRISS,
  long = Near InfraRed Imager and Slitless Spectrograph,
  cite = {2023niriss}
}

\DeclareAcronym{ers}{
  short = ERS,
  long = Early Release Science
}

\DeclareAcronym{bfe}{
  short = BFE,
  long = Brighter-Fatter Effect,
  cite = {Antilogus2014,Hirata2020,Argyriou2023}
}

\DeclareAcronym{amigo}{
  short = \texttt{Amigo},
  long = Aperture Masking Interferometry Generative Observations
}

\DeclareAcronym{mle}{
  short = MLE,
  long = Maximum Likelihood Estimate
}

\DeclareAcronym{mast}{
  short = MAST,
  long = the Barbara A. Mikulski Archive for Space Telescopes
}

\DeclareAcronym{rml}{
  short = RML,
  long = Regularized Maximum Likelihood
}

\DeclareAcronym{mcmc}{
  short = MCMC,
  long = Monte Carlo Markov Chain
}

\DeclareAcronym{hmc}{
  short = HMC,
  long = Hamiltonian Monte Carlo
}

\DeclareAcronym{nuts}{
  short = NUTS,
  long = No U-Turn Sampler,
  cite = {numpyro, cabezas2024blackjax, abril2023pymc, stan2023manual}
}

\DeclareAcronym{autodiff}{
  short = autodiff,
  long = automatic differentiation
}

\DeclareAcronym{gpu}{
  short = GPU,
  long = Graphics Processing Unit
}

\DeclareAcronym{disco}{
  short = DISCO,
  long = Delay-Insensitive Subspace of Calibrated Observables,  
  short-plural = s,
  long-plural  = s
}

\DeclareAcronym{opd}{
  short = OPD,
  long = Optical Path Difference
}

\DeclareAcronym{cmos}{
  short = CMOS,
  long = Complementary Metal-Oxide-Semiconductor
}

\DeclareAcronym{wr137}{
    short=WR~137,
    long={Wolf-Rayet 137}
}

\DeclareAcronym{lasso}{
    short=LASSO,
    long={Least Absolute Shrinkage and Selection Operator},
    cite={lasso}
}

\DeclareAcronym{smbh}{
    short=SMBH,
    long={supermassive black hole},
}

\DeclareAcronym{lbti}{
    short=LBTI,
    long={Large Binocular Telescope Interferometer},
    cite = {lbti}
}

\DeclareAcronym{stsci}{
    short=STScI,
    long={the Space Telescope Science Institute},
}

%% file: WR137SCHEDULE.tex
\begin{table}[ht!]
\centering
\caption{The JWST/AMI observing schedule for WR\,137 and its calibrator star (HD\,228337). Each row represents an individual exposure. All dates are in 2022, when mirror tilt events were more common as the JWST mirror segments were still settling. The horizontal dashed line between July 13th and July 15th indicated the time at which the mirror tilt event occurred. The dither column refers to the dither position within the primary dither pattern, with the stare remaining in the centre of the sub-array. The number of groups and integrations of the exposures are given by $n_\text{g}$ and $n_\text{int}$ respectively. The full integration time of an exposure is given by $t$. Roll is the roll angle of the telescope with respect to north.}
\label{tab:wr137_schedule}
\begin{tabular}{cccccccc}
Date                                               & Star                                             & Filter                                          & Dither                    & $n_{\text{g}}$                              & $n_{\text{int}}$                              & $t$ {[}min{]}                                 & Roll [\textdegree]                               \\ \hline
\cellcolor[HTML]{FAE8E0}                           & \cellcolor[HTML]{BBE7FE}                         & \cellcolor[HTML]{FFC2C7}                        & 1                         & \cellcolor[HTML]{FCFCFC}                    &                                               & \cellcolor[HTML]{FEF2EE}                      & \cellcolor[HTML]{FAE8E0}                        \\
\cellcolor[HTML]{FAE8E0}                           & \cellcolor[HTML]{BBE7FE}                         & \cellcolor[HTML]{FFC2C7}                        & 2                         & \cellcolor[HTML]{FCFCFC}                    &                                               & \cellcolor[HTML]{FEF2EE}                      & \cellcolor[HTML]{FAE8E0}                        \\
\cellcolor[HTML]{FAE8E0}                           & \cellcolor[HTML]{BBE7FE}                         & \cellcolor[HTML]{FFC2C7}                        & 3                         & \cellcolor[HTML]{FCFCFC}                    &                                               & \cellcolor[HTML]{FEF2EE}                      & \cellcolor[HTML]{FAE8E0}                        \\
\cellcolor[HTML]{FAE8E0}                           & \cellcolor[HTML]{BBE7FE}                         & \multirow{-4}{*}{\cellcolor[HTML]{FFC2C7}F480M} & 4                         & \multirow{-4}{*}{\cellcolor[HTML]{FCFCFC}4} & \multirow{-4}{*}{400}                         & \multirow{-4}{*}{\cellcolor[HTML]{FEF2EE}2.0} & \cellcolor[HTML]{FAE8E0}                        \\
\cellcolor[HTML]{FAE8E0}                           & \cellcolor[HTML]{BBE7FE}                         & \cellcolor[HTML]{8FDDE7}                        & \cellcolor[HTML]{FCFCFC}1 &                                             & \cellcolor[HTML]{FCFCFC}                      & \cellcolor[HTML]{FFF8F5}                      & \cellcolor[HTML]{FAE8E0}                        \\
\cellcolor[HTML]{FAE8E0}                           & \cellcolor[HTML]{BBE7FE}                         & \cellcolor[HTML]{8FDDE7}                        & \cellcolor[HTML]{FCFCFC}2 &                                             & \cellcolor[HTML]{FCFCFC}                      & \cellcolor[HTML]{FFF8F5}                      & \cellcolor[HTML]{FAE8E0}                        \\
\multirow{-7}{*}{\cellcolor[HTML]{FAE8E0}Jul 13} & \multirow{-7}{*}{\cellcolor[HTML]{BBE7FE}WR137} & \multirow{-3}{*}{\cellcolor[HTML]{8FDDE7}F380M} & \cellcolor[HTML]{FCFCFC}4 & \multirow{-3}{*}{2}                         & \multirow{-3}{*}{\cellcolor[HTML]{FCFCFC}680} & \multirow{-3}{*}{\cellcolor[HTML]{FFF8F5}1.7} & \multirow{-7}{*}{\cellcolor[HTML]{FAE8E0}192.0} \\ \hdashline
\cellcolor[HTML]{D8A7B1}                           & \cellcolor[HTML]{D3B5E5}                         & \cellcolor[HTML]{FFC2C7}F480M                   &                           & \cellcolor[HTML]{FCFCFC}7                   & 1020                                          & \cellcolor[HTML]{F882AA}9.0                   &                                                 \\
\cellcolor[HTML]{D8A7B1}                           & \multirow{-2}{*}{\cellcolor[HTML]{D3B5E5}CAL}    & \cellcolor[HTML]{8FDDE7}F380M                   &                           & \cellcolor[HTML]{FCFCFC}2                   & 2720                                          & \cellcolor[HTML]{FCCBC7}6.8                   & \multirow{-2}{*}{}                              \\
\cellcolor[HTML]{D8A7B1}                           & \cellcolor[HTML]{BBE7FE}                         & \cellcolor[HTML]{FFC2C7}F480M                   &                           & \cellcolor[HTML]{FCFCFC}4                   & 1600                                          & \cellcolor[HTML]{FAA8B7}8.0                   & \cellcolor[HTML]{D8A7B1}                        \\
\multirow{-4}{*}{\cellcolor[HTML]{D8A7B1}Jul 15} & \multirow{-2}{*}{\cellcolor[HTML]{BBE7FE}WR137} & \cellcolor[HTML]{8FDDE7}F380M                   & \multirow{-4}{*}{Stare}   & \cellcolor[HTML]{FCFCFC}2                   & 2720                                          & \cellcolor[HTML]{FCCBC7}6.8                   & \multirow{-2}{*}{\cellcolor[HTML]{D8A7B1}189.8} \\ \hline
\cellcolor[HTML]{B6E2D3}                           & \cellcolor[HTML]{BBE7FE}                         & \cellcolor[HTML]{FFC2C7}                        & \cellcolor[HTML]{FCFCFC}1 &                                             & \cellcolor[HTML]{FCFCFC}                      & \cellcolor[HTML]{FEF2EE}                      & \cellcolor[HTML]{B6E2D3}                        \\
\cellcolor[HTML]{B6E2D3}                           & \cellcolor[HTML]{BBE7FE}                         & \cellcolor[HTML]{FFC2C7}                        & \cellcolor[HTML]{FCFCFC}2 &                                             & \cellcolor[HTML]{FCFCFC}                      & \cellcolor[HTML]{FEF2EE}                      & \cellcolor[HTML]{B6E2D3}                        \\
\cellcolor[HTML]{B6E2D3}                           & \cellcolor[HTML]{BBE7FE}                         & \cellcolor[HTML]{FFC2C7}                        & \cellcolor[HTML]{FCFCFC}3 &                                             & \cellcolor[HTML]{FCFCFC}                      & \cellcolor[HTML]{FEF2EE}                      & \cellcolor[HTML]{B6E2D3}                        \\
\cellcolor[HTML]{B6E2D3}                           & \cellcolor[HTML]{BBE7FE}                         & \multirow{-4}{*}{\cellcolor[HTML]{FFC2C7}F480M} & \cellcolor[HTML]{FCFCFC}4 & \multirow{-4}{*}{4}                         & \multirow{-4}{*}{\cellcolor[HTML]{FCFCFC}400} & \multirow{-4}{*}{\cellcolor[HTML]{FEF2EE}2.0} & \cellcolor[HTML]{B6E2D3}                        \\
\cellcolor[HTML]{B6E2D3}                           & \cellcolor[HTML]{BBE7FE}                         & \cellcolor[HTML]{8FDDE7}                        & 1                         & \cellcolor[HTML]{FCFCFC}                    &                                               & \cellcolor[HTML]{FFF8F5}                      & \cellcolor[HTML]{B6E2D3}                        \\
\cellcolor[HTML]{B6E2D3}                           & \cellcolor[HTML]{BBE7FE}                         & \cellcolor[HTML]{8FDDE7}                        & 2                         & \cellcolor[HTML]{FCFCFC}                    &                                               & \cellcolor[HTML]{FFF8F5}                      & \cellcolor[HTML]{B6E2D3}                        \\
\cellcolor[HTML]{B6E2D3}                           & \cellcolor[HTML]{BBE7FE}                         & \cellcolor[HTML]{8FDDE7}                        & 3                         & \cellcolor[HTML]{FCFCFC}                    &                                               & \cellcolor[HTML]{FFF8F5}                      & \cellcolor[HTML]{B6E2D3}                        \\
\cellcolor[HTML]{B6E2D3}                           & \multirow{-8}{*}{\cellcolor[HTML]{BBE7FE}WR137} & \multirow{-4}{*}{\cellcolor[HTML]{8FDDE7}F380M} & 4                         & \multirow{-4}{*}{\cellcolor[HTML]{FCFCFC}2} & \multirow{-4}{*}{680}                         & \multirow{-4}{*}{\cellcolor[HTML]{FFF8F5}1.7} & \multirow{-8}{*}{\cellcolor[HTML]{B6E2D3}162.7} \\
\cellcolor[HTML]{B6E2D3}                           & \cellcolor[HTML]{D3B5E5}                         & \cellcolor[HTML]{FFC2C7}                        & \cellcolor[HTML]{FCFCFC}1 &                                             & \cellcolor[HTML]{FCFCFC}                      & \cellcolor[HTML]{FDEBE7}                      &                                                 \\
\cellcolor[HTML]{B6E2D3}                           & \cellcolor[HTML]{D3B5E5}                         & \cellcolor[HTML]{FFC2C7}                        & \cellcolor[HTML]{FCFCFC}2 &                                             & \cellcolor[HTML]{FCFCFC}                      & \cellcolor[HTML]{FDEBE7}                      &                                                 \\
\cellcolor[HTML]{B6E2D3}                           & \cellcolor[HTML]{D3B5E5}                         & \cellcolor[HTML]{FFC2C7}                        & \cellcolor[HTML]{FCFCFC}3 &                                             & \cellcolor[HTML]{FCFCFC}                      & \cellcolor[HTML]{FDEBE7}                      &                                                 \\
\cellcolor[HTML]{B6E2D3}                           & \cellcolor[HTML]{D3B5E5}                         & \multirow{-4}{*}{\cellcolor[HTML]{FFC2C7}F480M} & \cellcolor[HTML]{FCFCFC}4 & \multirow{-4}{*}{7}                         & \multirow{-4}{*}{\cellcolor[HTML]{FCFCFC}255} & \multirow{-4}{*}{\cellcolor[HTML]{FDEBE7}2.2} &                                                 \\
\cellcolor[HTML]{B6E2D3}                           & \cellcolor[HTML]{D3B5E5}                         & \cellcolor[HTML]{8FDDE7}                        & 1                         & \cellcolor[HTML]{FCFCFC}                    &                                               & \cellcolor[HTML]{FFF8F5}                      &                                                 \\
\cellcolor[HTML]{B6E2D3}                           & \cellcolor[HTML]{D3B5E5}                         & \cellcolor[HTML]{8FDDE7}                        & 2                         & \cellcolor[HTML]{FCFCFC}                    &                                               & \cellcolor[HTML]{FFF8F5}                      &                                                 \\
\cellcolor[HTML]{B6E2D3}                           & \cellcolor[HTML]{D3B5E5}                         & \cellcolor[HTML]{8FDDE7}                        & 3                         & \cellcolor[HTML]{FCFCFC}                    &                                               & \cellcolor[HTML]{FFF8F5}                      &                                                 \\
\multirow{-16}{*}{\cellcolor[HTML]{B6E2D3}Aug 8} & \multirow{-8}{*}{\cellcolor[HTML]{D3B5E5}CAL}    & \multirow{-4}{*}{\cellcolor[HTML]{8FDDE7}F380M} & 4                         & \multirow{-4}{*}{\cellcolor[HTML]{FCFCFC}2} & \multirow{-4}{*}{680}                         & \multirow{-4}{*}{\cellcolor[HTML]{FFF8F5}1.7} & \multirow{-8}{*}{}                             
\end{tabular}
\end{table}